\documentclass[conference]{IEEEtran}
\usepackage{mathtools, cuted} 
\usepackage{lettrine} 
\usepackage{amsthm,amsmath} 
\usepackage{epsfig,amssymb,amsbsy,verbatim,subfigure,array}
\usepackage{pstricks,cite,url}%caption
\usepackage{multicol,afterpage,wrapfig,paralist} %fancyheadings,
\usepackage{hyperref}
\usepackage{breakurl}
\usepackage{verbatim,tikz,subfigure}
\usepackage{tikz,pgfplots}
\usepackage[english]{babel}
 
{}

\def\d{\mathrm{d}}
\title{Analyzing Optical TDMA to Mitigate Interference in Downlink LiFi Optical Attocell Networks}
\author{\IEEEauthorblockN{Atchutananda Surampudi}
\IEEEauthorblockA{Department of Engineering Science, University of Oxford, OX1-3PJ, Oxford. \\
atchutananda.surampudi@wadham.ox.ac.uk}
}
\begin{document}
\maketitle

\begin{abstract}
Co-channel interference in the downlink of LiFi attocell networks significantly decreases the network performance in terms of rate. Analysis of multiple access schemes is essential to mitigate interference and improve rate. The light emitting diodes (LEDs) being centrally monitored, the time division multiple access (TDMA) scheme over the LEDs will be suitable to analyze. This work considers the interference characterization in [1] over M-PAM modulated signals to derive an exact expression for the goodput $G$ of the time scheduled attocell network, which is arranged as a deterministic square lattice in two dimensions. Given this TDMA over the LEDs, numerical simulations show that the LEDs can be optimally time scheduled to maximize the goodput, which implies that the TDMA mitigates interference in an attocell network compared to the case when the LEDs are unscheduled.
\end{abstract}

\section{Introduction}
Light Fidelity (LiFi) has emerged as a high speed wireless data access solution using visible light [2]. The light emitting diode (LED) is used as an access point which both illuminates a certain area and provides a wireless connection in that area of illumination. The arrangement of LEDs as a network is called an attocell network. Such an attocell network is centrally monitored and is generally arranged in a deterministic lattice in both one and two dimensions. For downlink access, the signal-to-interference-plus-noise-ratio (SINR) is analyzed at a receiver to measure the system performance. Co-channel interference or simply interference in such networks significantly affects the SINR and hence limits the data rate. Tractable closed form expressions characterizing this interference have been provided in [1] over the attocell geometries. While various multiple access schemes have been used  to mitigate interference in conventional wireless networks, the time division multiple access (TDMA) over the LiFi LEDs will be suitable to analyze and implement since the attocell networks are always centrally monitored over a deterministic lattice.
\subsection{Related works}
To the best of the authors' knowledge, due to the absence of tractable closed form expressions for interference and hence the SINR in LiFi attocell networks, the topic of time scheduling to mitigate interference has relatively been untouched. Nonetheless, there have been other approaches to achieve mitigation [3-7], respectively wherein the research focussed on angle diversity receivers and their orientation and some novel resource allocation schemes. All the mentioned works, however significant they actually are, do not consider TDMA over the LEDs to improve the network performance; which actually is an important aspect since the entire attocell network tends to be centrally monitored.   
\subsection{Contributions of this work} 
\begin{itemize}
\item Since the LiFi LEDs are centrally monitored, the TDMA is implemented over the LEDs to mitigate interference over a two dimensional LiFi attocell network. 
\item Using the results in [1] over M-PAM modulated signals, an exact expression  \eqref{eqn:good} is derived an for the goodput $G$ of the TDMA system, where $G$ is defined as the product of the rate and the probability of correctness of reception.
\item The existence of an optimum TDMA parameter $K$ is shown, at which $G$ is maximized. 
\end{itemize}
\subsection{Arrangement of the paper and notations}
The involved notations are given in Table I. 
\begin{table}
\label{t/data1}
\caption{Notations used in the paper}
\label{abc}
\begin{tabular}{ | m{1cm} | m{1.9cm}| m{4cm} |} 
\hline
S.No.& NOTATION & MEANING\\ 
\hline
1 & $z=\sqrt{z_{x}^{2}+z_{y}^{2}}$ & Distance of the PD from the origin. The location is given by Cartesian coordinates $(z_{x},z_{y})$, where $z_x$ and $z_y$ are measured in metres.\\
\hline
2 & $h$ & Height of LED installation.\\
\hline
%3 & $h_{1}$ & $\sqrt{h^{2}+z_{y}^{2}}$, used in the 1D model.\\
%\hline
3 & $m, \beta, \theta_{h}$ & Terms representing the HPSA of the LED.\\
\hline
4 & $a$ & Square lattice edge length or inter LED spacing.\\
\hline
5 & $A_{pd}$ & Area of receiver PD.\\
\hline
6 & $R_{pd}$ & Responsivity of PD.\\
\hline
7 & $(i,j)$  & Indices representing an LED in the network.\\
\hline
8 & $D_{i,j}$ & The distance, on ground, between the PD and an $(i,j)^{th}$ interfering LED.\\
\hline
9 & $\theta_{f}$ & Field-of-View (FOV) of the PD.\\
\hline
10 & $G_{i,j}(z)$ & Channel gain from an $(i,j)^{th}$ LED to a PD at $z$ from origin.\\
\hline
%11 & $\rho(D_{i})$ & FOV constraint function as defined in the paper.\\
%\hline
%13 & $i, (u,v)$ & Indices for the interfering LEDs in one and two dimension models respectively.\\
%\hline
11 & $\gamma(z)$ & The SINR at $z$.\\
\hline
12 & $x_{i,j}(t)$ & Baseband signal from the $(i,j)^{th}$ LED. \\
\hline
13 & $s_{i,j}(t)$ & Intensity modulated signal from $x_{i,j}(t)$.\\
\hline
%17 & $r_{o}$ & Equivalent internal resistance of the PD.\\
%\hline
%18 & $\mathbb{I}_{\infty}(z)$ & Normalized Interference due to infinite number of interferers. Used in both 1D and 2D networks.\\
%\hline
%19 & $\mathbb{I}_{n}(z)$ & Normalized Interference due to finite number $(n)$ of interferers. Used in both 1D and 2D networks.\\
%\hline
14 & $\hat{\mathtt{I}}_{u,v}(z)$ & Series approximation for the variance $\sigma_{1}^{2}$ over $(u,v)$ terms.\\
\hline
15 & $\hat{\mathcal{I}}_{u,v}(z)$ & Series approximation for the mean $\mu$ over $(u,v)$ terms.\\
\hline
16 & $(u,v)$ & Number of approximation terms. \\
\hline
%21 & $\iota$ & Imaginary number $\sqrt{-1}$.\\
%\hline
%17 & $k$ & Number of approximation terms for 1D model. \\
%\hline
%18 & $w$ & Frequency index in 1D model. \\
%\hline
17 & $(w,f)$ & Fourier index. \\
\hline
%26 & $e(k)$ or $\xi(j,l)$ & The approximation error in 1D and 2D models respectively.\\
%\hline
%27 & $\hat{e}$ or $\hat{\xi}$ & Percentage approximation error in 1D and 2D models.\\
%\hline
%28 & $O(.)$ or $\Theta(.)$ & Asymptotic notation as defined in the paper.\\
%\hline
%1 & $$ & y.\\
%\hline
\end{tabular}
\end{table}
%\label{parameter}
The paper is arranged as follows. Section II describes the system model. Section III describes the time scheduling scheme along with the numerical simulations validating the analytical results. The paper concludes with Section IV.         

\section{System model}
\subsection{The attocell network}
\begin{figure}[ht]  
\centering
 \resizebox{0.5\textwidth}{!}{%
 \begin{tikzpicture}
 \draw (0,0) -- (0,3) -- (5,8) -- (14,8) -- (14,5);
 \draw (14,5) -- (9,0) -- (0,0);
 \draw (14,8) -- (9,3) -- (0,3);
 \draw (9,0) -- (9,3);
 \draw [dotted] (0,0) -- (5,5) -- (14,5);
 \draw [dotted] (5,5) -- (5,8);
 
 \draw [line width= 0.03cm,->] (0,0) -- (1.6,0);
 \node [below] at (1.6,0) {\large $x$};
 %\draw [line width= 0.03cm,->] (0,0) -- (0,1.3);
 \draw [line width= 0.03cm,->] (0,0) -- (1.3,1.3);
  \node [below] at (1.3,1.3) {\large $y$};
 
 \draw [fill] (0.5,3.5) circle [radius=0.10];
 %\node [below] at (0.5,3.5) {\scriptsize $(-2a,-a,h)$};
 
 \draw [fill] (2.5,3.5) circle [radius=0.10];
 %\node [below] at (2.5,3.5) {\scriptsize $(-a,-a,h)$};
 
 \draw [fill] (4.5,3.5) circle [radius=0.10];
 \node [below] at (4.5,3.5) {\large $(0,-a,h)$};
 
 %\draw [fill] (6.5,3.5) circle [radius=0.10];
 %\node [below] at (6.5,3.5) {\scriptsize $(a,-a,h)$};
 
 %\draw [fill] (8.5,3.5) circle [radius=0.10];
 %\node [below] at (8.5,3.5) {\scriptsize $(2a,-a,h)$};
 
 \draw [fill] (2.5,5.5) circle [radius=0.10];
 \node [below] at (2.5,5.5) {\large $(-2a,0,h)$};
 \node at (5.5,1.8) {\large Attocell};
 
 \draw [fill] (4.5,5.5) circle [radius=0.10];
 \node [below] at (4.5,5.5) {\large $(-a,0,h)$};
 
 \draw [fill] (6.5,5.5) circle [radius=0.10];
 \node [below] at (6.5,5.5) {\large $(0,0,h)$};
 \node [above] at (6.5,5.5) {\large tagged-LED};
 
 \draw [fill] (8.5,5.5) circle [radius=0.10];
 \node [below] at (8.5,5.5) {\large $(a,0,h)$};
 
 \draw [fill] (10.5,5.5) circle [radius=0.10];
 \node [below] at (10.5,5.5) {\large $(2a,0,h)$};
 
 \draw [fill] (4.5,7.5) circle [radius=0.10];
 %\node [below] at (4.5,7.5) {\scriptsize $(-2a,a,h)$};
 
 \draw [fill] (6.5,7.5) circle [radius=0.10];
 %\node [below] at (6.5,7.5) {\scriptsize $(-a,a,h)$};
 
 \draw [fill] (8.5,7.5) circle [radius=0.10];
 \node [below] at (8.5,7.5) {\large $(0,a,h)$};
 
 \draw [fill] (10.5,7.5) circle [radius=0.10];
 %\node [below] at (10.5,7.5) {\scriptsize $(a,a,h)$};
 
 \draw [fill] (12.5,7.5) circle [radius=0.10]; 
 %\node [below] at (12.5,7.5) {\scriptsize $(2a,a,h)$};
 
 \draw [dotted] (1,0) -- (6,5);
 \draw [dotted] (3,0) -- (8,5);
 \draw [dotted] (5,0) -- (10,5);
 \draw [dotted] (7,0) -- (12,5);
 \draw [dotted] (1.5,1.5) -- (10.5,1.5);
 \draw [dotted] (3.5,3.5) -- (12.5,3.5);
 \draw [dotted] (6.5,2.5) -- (6.5,5.5);
 \node [align=center] at (6.5,2.5) {\small x};
 \node [below] at (6.5,2.5) {\large $(0,0,0)$};
 \node [left] at (6.5,4.5) {\large $h$}; 
 \draw (7.35,2.99) -- (7.55,2.99) -- (7.65,3.09) -- (7.45,3.09) -- (7.35,2.99);
 \draw [fill] (7.35,3.09) -- (7.55,3.09) -- (7.65,3.19) -- (7.45,3.19) -- (7.35,3.09);
 \draw (7.35,2.99) -- (7.35,3.09);
 \draw (7.55,2.99) -- (7.55,3.09);
 \draw (7.65,3.09) -- (7.65,3.19);
 \draw (7.45,3.09) -- (7.45,3.19);
 \draw [dotted] (6.5,2.5) -- (7.5,3.04);
 %\node [below] at (8.3,3.04) {\large $(d_{x},d_{y},0)$}; 
\node [below] at (8.4,4.04) {\large PD at $(z_{x},z_{y},0)$};
 \node [below] at (7,2.77) {\large $z$};
 \draw [dashed] (6.5,5.5) -- (7.5,3.04); 
 \draw [dashed] (6.5,1.5) -- (8.5,3.5) -- (6.5,3.5) -- (4.5,1.5) -- (6.5,1.5);
 \node [above] at (5.5,2.5) {\large $a$};
 \node [below] at (5.5,1.5) {\large $a$};
 \node [above] at (9,8) {\large Infinite two dimension plane}; 
 
 \end{tikzpicture} 
}%
\caption{This figure, taken from \cite{atchu2}, shows the infinite two dimensional model. There are infinite number of LEDs (circular dots) indexed as $(i,j)\in\mathbb{S}$ and arranged symmetrically at regular intervals of $a$ as a uniform square grid, all over the plane, installed at a height $h$. The rectangular dotted regions on ground depict the attocells corresponding to each $(i,j)^{th}$ LED above. The user PD (small cuboid) at $(z_{x},z_{y},0)$ (inside one of the attocell), receives data wirelessly from the tagged-LED corresponding to the attocell in which it is located. Here, that attocell is highlighted as dash-dot. All other LEDs are co-channel interferers. Here we assume that the user PD can move anywhere on the ground plane. For a finite network, the interferers are assumed to be located symmetrically around the tagged LED.}
\label{two_dim}
\end{figure}
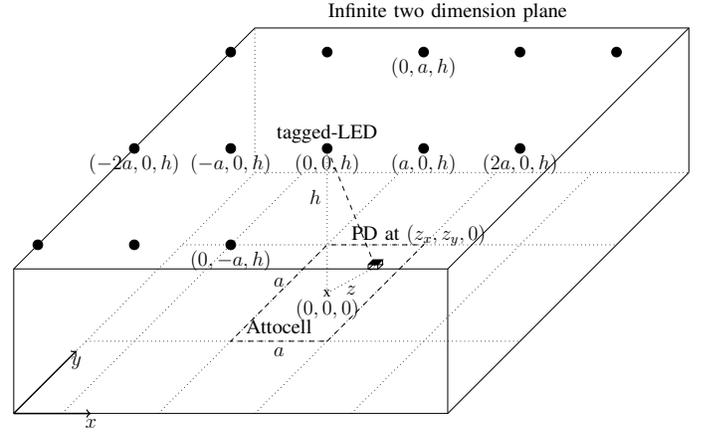

Consider the two dimensional LiFi attocell network in Fig.\ref{two_dim}. Let the set $\mathbb{S}$ represent an infinite set of LiFi LEDs with $(i,j)\in\mathbb{S}$ indicating the two dimensional index of a particular LiFi LED. All the LEDs are fixed at a height $h$, are separated symmetrically by a distance $a$ and emit at a uniform average optical power $P_{o}$. The photodiode (PD) is assumed to have it's surface always parallel to the ground, i.e., without any orientation towards any LED and is assumed to be located at $(z_{x},z_{y},0)$ from the origin. We neglect any non-linearities of the LED while intensity modulation and assume the field-of-view (FOV) $\theta_{f}$ of the PD to be $\frac{\pi}{2}$ radians. The LEDs have a lambertian emission order $m$ given as $m = -\frac{\ln(2)}{\ln(\cos(\theta_{h}))}$, where $\theta_{h}$ is the half-power-semi-angle (HPSA) of any given LED. Let the PD have a cross section area $A_{pd}$ and the responsivity $R_{pd}$. Let the optical system bandwidth be $W$, and the noise power spectral density at the PD be $N_{o}$. Also, since the entire network is considered to be located in an open area without any opaque obstructions, the works in \cite{nlos1,nlos2,nlos3} will be followed to neglect any multipath and non-line-of-sight components received at the PD. Importantly, this PD tags to the nearest LED, which is considered to be at $(0,0,h)$.

\subsection{The modulation scheme} 
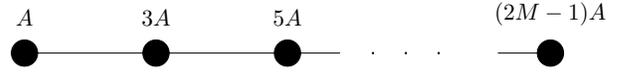
\begin{figure}[ht]
\centering
\resizebox{0.45\textwidth}{!}{%
 \begin{tikzpicture} 
 \draw [line width= 0.01cm] (0.2,0.5) -- (5,0.5);
 \draw [line width= 0.01cm] (7.4,0.5) -- (8.2,0.5);
 \draw [fill] (0.2,0.5) circle [radius=0.20];
 \node [above] at (0.2,0.8) {\normalsize $A$};
 \draw [fill] (2.2,0.5) circle [radius=0.20];
 \node [above] at (2.2,0.8) {\normalsize $3A$};
 \draw [fill] (4.2,0.5) circle [radius=0.20];
 \node [above] at (4.2,0.8) {\normalsize $5A$};
 \draw [fill] (5.5,0.5) circle [radius=0.01];
 \node [above] at (8.2,0.8) {\normalsize $(2M-1)A$};
 \draw [fill] (6,0.5) circle [radius=0.01];
 \draw [fill] (6.5,0.5) circle [radius=0.01];
 \draw [fill] (8.2,0.5) circle [radius=0.20];
 \end{tikzpicture}
 }%
\caption{Here, the \textit{M}-PAM constellation of the transmitted intensity at any LED is shown. Each possible point in the constellation is assumed to be equiprobably transmitted with a probability $\frac{1}{M}$. $A$ is a constant in Watts.}
\label{pam} 
\end{figure}

In LiFi, the intensity modulation and direct detection (IM/DD) schemes are used to form a wireless link \cite{islim}. The data frames $x_{i,j}(t)$ from every $(i,j)^{th}$ LED are intensity modulated to $s_{i,j}(t)$ over the visible light intensities. The single carrier \textit{M}-ary pulse-amplitude-modulation (PAM) scheme is assumed with equiprobable intensity levels; each level with probability $\frac{1}{M}$, and ranging from $A$ to $(2M-1)A$, where $A$ is a constant in watts, as shown in Fig. \ref{pam}. 

It is assumed that $\log_{2}(M)$ bits of information are transmitted over a given time slot from a particular $(i,j)^{th}$ LED. The average optical power $P_{o}$ emitted from this $(i,j)^{th}$ LED is given as
\begin{align*} 
P_{o}&=\mathbb{E}(s_{i,j}(t)),\nonumber\\
&=\frac{A}{M}(1+3+...+(2M-1)),\nonumber\\
&=AM,
\label{eqn:pam}
\end{align*}
where $\mathbb{E}(.)$ is the expectation operator over the possible intensity values. So, from \cite[Eqn. 2]{atchu2}, the channel gain $G_{i,j}(z)$ experienced by the optical intensities from the $(i,j)^{th}$ LED in line of sight with the PD is given as 
\begin{equation*}  
G_{i,j}(z) = \frac{(m+1)A_{pd}h^{m+1}}{2\pi}(D_{i,j}^{2}+h^{2} )^{\frac{-(m+3)}{2}},
\label{eqn:gain2}   
\end{equation*}
where $D_{i,j}=\sqrt{(z_{x}+ia)^{2}+(z_{y}+ja)^{2}}$, represents the horizontal distance between the PD and the $i^{th}$ LED at $(ia,ja,h)$. All the LEDs, saving the tagged LED at $(0,0,h)$, are termed as co-channel interferers. Now, the total received current $I(z,t)$ at the PD, during a time slot $t$ is given as 
\[ I(z,t) =s_{0}(t) G_{0}(z) R_{pd} + \mathcal{I}_{\infty}(z) + n(t),\]
where $\mathcal{I}_{\infty}(z)=\sum_{(i,j) \in \mathbb{S}\setminus (0,0)}s_{i,j}(t)G_{i,j}(z)R_{pd}$ is the co-channel interference current, and $n(t)\sim\mathcal{N}(0,\sigma^{2})$ has a power spectral density $N_{o}$ such that the variance of the noise process $\sigma^{2}=N_{o}W$. Also, the co-channel interference term $s_{i,j}(t) G_{i,j}(z) R_{pd}$, is a uniformly distributed random variable over the \textit{M}-PAM intensity levels whose mean is given as    
\begin{align*}
I_{i,j}(z)&=\mathbb{E}[s_{i,j}(t) G_{i,j}(z) R_{pd}],\nonumber\\
&=AMG_{i,j}(z)R_{pd},
\end{align*}
and the variance is given as $\frac{A^{2}(M^{2}-1)}{3}G_{i,j}^{2}(z) R_{pd}^{2}$. Summing up over a large number of interfering LEDs, the sum $\sum_{i,j \in \mathbb{S}\setminus 0}s_{i,j}(t) G_{i,j}(z) R_{pd}$ is approximated to converge in distribution to a Gaussian random variable with a mean of 
\begin{align}
\mu&=\mathbb{E}[\mathcal{I}_{\infty}(z)],\nonumber\\
%&=\sum_{i,j \in \mathbb{S}\setminus 0}\mathbb{E}[s_{i}(t) G_{i}(z) R_{pd}],\nonumber\\
&=AM\sum_{(i,j) \in \mathbb{S}\setminus (0,0)}G_{i,j}(z)R_{pd},\nonumber\\
&= \sum_{(i,j) \in \mathbb{S}\setminus (0,0)}T_{1}(D_{i,j}^{2} + h^{2} )^{\frac{-(m+3)}{2}},
\label{eqn:int1}
\end{align} 
and a variance of 
\begin{align}
\sigma_{1}^{2}&= \frac{A^{2}(M^{2}-1)}{3}\sum_{(i,j) \in \mathbb{S}\setminus (0,0)}G_{i,j}^{2}(z) R_{pd}^{2},\nonumber\\
&=\sum_{(i,j) \in \mathbb{S}\setminus (0,0)} T_{2}(D_{i,j}^{2} + h^{2} )^{-(m+3)},
\label{eqn:var}  
\end{align}
where $T_{1}=\frac{AMR_{pd}A_{pd}(m+1)h^{m+1}}{2\pi}$ and $T_{2}=\frac{A^{2}(M^{2}-1)R_{pd}^{2}A^{2}_{pd}(m+1)^{2}h^{2(m+1)}}{12\pi^{2}}$. So, the received interference plus noise $n(t)$ at the PD is normally distributed as $\mathcal{N}(\mu,\sigma^{2}+\sigma_{1}^{2})$. Using the interference characterization in \cite{atchu2} by considering till $(u,v)$ Fourier terms, we give a closed form expression for mean $\mu$ in \eqref{eqn:int1} as 
\begin{align} 
\mu & \approx \hat {\mathcal{I}}_{u,v}(z), \nonumber\\
&=T_{1} \bigg( \frac{h^{2-\beta}\pi}{a^{2}\big(\frac{\beta}{2}-1\big)}-\frac{1}{(z^{2}+h^{2})^{\frac{\beta}{2}}}+ \sum_{(w,f)\in\mathbb{A}} g_{1}(w,f)\bigg),
\label{eqn:intapp1} 
\end{align} 
where $\mathbb{A} \triangleq \{\mathbb{Z}^{2}\cap([0,u]\times[0,v])\}\setminus(0,0)$ over the set of integers $\mathbb{Z}^{2}, u$ and $v$; $\beta=m+3$ and $\Gamma(x)=\int_{0}^{\infty}t^{x-1}e^{-t}\d t$ is the standard gamma function. Also, 
\begin{align*}
g_{1}(w,f)=\frac{\mathbb{K}_{\frac{\beta}{2}-1}\bigg(\frac{2\pi h\sqrt{f^{2}+w^{2}}}{a}\bigg)\cos\big(\frac{2\pi wz_{x}}{a}\big)\cos\big(\frac{2\pi fz_{y}}{a}\big)}{\bigg(\frac{h}{2\pi\sqrt{f^{2}+w^{2}}}\bigg)^{\frac{\beta}{2}-1}2^{\frac{\beta}{2}-4} a^{\frac{\beta}{2}+1}\frac{\Gamma\big(\frac{\beta}{2}\big)}{\pi}},
\end{align*}
where $\mathbb{K}_{\tau}(y)=\frac{\Gamma(\tau+\frac{1}{2})(2y^{\tau})}{\sqrt{\pi}}\int_{0}^{\infty}\frac{\cos(t)\d t}{(t^{2}+y^{2})^{v+\frac{1}{2}}}$, is the modified Bessel function of second kind.       

Similarly, $\sigma_{1}^{2}$ in \eqref{eqn:var} can be approximated to a closed form expression as 
\begin{align}
\sigma_{1}^{2} & \approx \hat {\mathtt{I}}_{u,v}(z), \nonumber\\
&=T_{2} \bigg(\frac{h^{2-2\beta}\pi}{a^{2}(\beta-1)}-\frac{1}{(z^{2}+h^{2})^{\beta}}+ \sum_{(w,f)\in\mathbb{A}} g_{2}(w,f)\bigg),
\label{eqn:intapp1a}
\end{align}    
where,
\begin{align*}
g_{2}(w,f)=\frac{\mathbb{K}_{\beta-1}\bigg(\frac{2\pi h\sqrt{f^{2}+w^{2}}}{a}\bigg)\cos\big(\frac{2\pi wz_{x}}{a}\big)\cos\big(\frac{2\pi fz_{y}}{a}\big)}{\bigg(\frac{h}{2\pi\sqrt{f^{2}+w^{2}}}\bigg)^{\beta-1}2^{\beta-4} a^{\beta+1}\frac{\Gamma(\beta)}{\pi}}.
\end{align*}
\subsection{The Probability of Error}
\begin{figure}[ht]
\centering
\resizebox{0.45\textwidth}{!}{%
 \begin{tikzpicture} 
 \draw [line width= 0.01cm] (0.2,0.5) -- (5,0.5);
 \draw [line width= 0.01cm] (7.4,0.5) -- (8.2,0.5);
 \draw [fill] (0.2,0.5) circle [radius=0.20];
 \node [above] at (0.2,0.8) {\normalsize $AG_{0,0}(z)$};
 \draw [fill] (2.2,0.5) circle [radius=0.20];
 \node [below] at (2.2,0.2) {\normalsize $3AG_{0,0}(z)$};
 \draw [fill] (4.2,0.5) circle [radius=0.20];
 \node [above] at (4.2,0.8) {\normalsize $5AG_{0,0}(z)$};
 \draw [fill] (5.5,0.5) circle [radius=0.01];
 \node [below] at (8.2,0.2) {\normalsize $(2M-1)AG_{0,0}(z)$};
 \draw [fill] (6,0.5) circle [radius=0.01];
 \draw [fill] (6.5,0.5) circle [radius=0.01];
 \draw [fill] (8.2,0.5) circle [radius=0.20];
 \draw [dotted] (1.2,0) -- (1.2,1.2);
 \draw [dotted] (3.2,0) -- (3.2,1.2);
 \draw [dotted] (2.2,0.5) -- (2.2,2);
 \draw [dotted] (4.2,0.5) -- (4.2,2);
 \node at (3.2,1.9) {\normalsize $d$};
 \draw[arrows=->](3.5,1.9)--(4.2,1.9);
 \draw[arrows=->](2.9,1.9)--(2.2,1.9);
 \end{tikzpicture}
 }%
\caption{Here, the \textit{M}-PAM constellation of the received intensity at the PD is shown when the effect of noise and interference is not considered. The intensities suffer through a channel gain $G_{0,0}(z)$.}
\label{pam2} 
\end{figure}
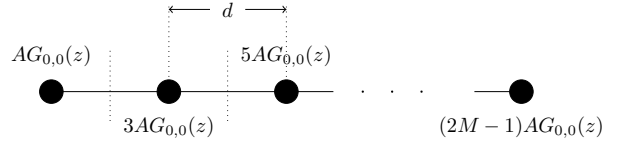

For an \textit{M}-PAM IM/DD scheme, the probability of error depends on two factors, namely, the distance $d$ between two adjacent constellation points of \textit{M}-PAM, and the interference plus noise at the PD. We neglect any non-linearities of the PD while reception. The tagged LED at $(0,0,h)$ transmits data to the PD over $M$ equiprobably different intensity levels. If there were no effect of interference or noise at the PD, then every $l^{th}$ intensity level would just suffer through a channel gain $G_{0,0}(z)$, and hence be received as $(2l-1)AG_{0,0}(z)$ at the PD. So, the adjacent distance $d$ between two constellation points would be $2AG_{0,0}(z)$, as shown in Fig. \ref{pam2}. \\ \par

If the noise and the interference were present at the PD, the received symbols would be in error. For a given constellation point, the per symbol probability of error $P_{s}$ can be calculated using the mean $(\mu)$ and variance$(\sigma_{1}^{2})$ respectively in \eqref{eqn:int1} and \eqref{eqn:var} as 
\begin{align*}
P_{s}&=\mathbb{P}\bigg[\mathcal{I}_{\infty}(z)+n(t)>\frac{R_{pd}d}{2}\bigg],\nonumber\\
%&\stackrel{(a)}{=}\frac{1}{\sigma \sqrt{2\pi}}\int_{\frac{R_{pd}d}{2}}^{\infty}e^{-\frac{(x-\mu)^{2}}{2(\sigma^{2}+\sigma_{1}^{2})}}\d x, \nonumber\\
%&=\frac{1}{\sqrt{2\pi}}\int_{\frac{\frac{R_{pd}d}{2}-\mu}{\sqrt{\sigma^{2}+\sigma_{1}^{2}}}}^{\infty}e^{-\frac{u^{2}}{2}}\d u, \nonumber\\
&=Q\Bigg(\frac{\frac{R_{pd}d}{2}-\mu}{\sqrt{\sigma^{2}+\sigma_{1}^{2}}}\Bigg),\nonumber\\
&\stackrel{(a)}{=}Q\Bigg(\frac{\frac{R_{pd}d}{2}-\hat{\mathcal{I}}_{u,v}(z)}{\sqrt{\sigma^{2}+\hat{\mathtt{I}}_{u,v}(z)}}\Bigg),
\end{align*}
where $(a)$ follows the approximation in \eqref{eqn:intapp1} and \eqref{eqn:intapp1a}. Assuming the leftmost and the rightmost constellation points have neighbours at infinity, the total symbol error $P_{e}$, using the union bound is given as 
\begin{align}
P_{e}&\leq \frac{2}{M} Q\Bigg(\frac{\frac{R_{pd}d}{2}-\hat{\mathcal{I}}_{u,v}(z)}{\sqrt{\sigma^{2}+\hat{\mathtt{I}}_{u,v}(z)}}\Bigg)+\frac{2}{M}\sum_{j=2}^{M-1}Q\Bigg(\frac{\frac{R_{pd}d}{2}-\hat{\mathcal{I}}_{u,v}(z)}{\sqrt{\sigma^{2}+\hat{\mathtt{I}}_{u,v}(z)}}\Bigg),\nonumber\\
&=\frac{2(M-1)}{M} Q\Bigg(\frac{\frac{R_{pd}d}{2}-\hat{\mathcal{I}}_{u,v}(z)}{\sqrt{\sigma^{2}+\hat{\mathtt{I}}_{u,v}(z)}}\Bigg).
\label{eqn:probe}
\end{align}

\subsection{The Rate Expression}
The SINR is calculated using the electrical powers received at the PD as 
\begin{align}
\gamma(z)&=\frac{T_{1}^{2}(z^{2}+h^{2})^{-m-3}}{\sigma_{1}^{2}+\sigma^{2}}, \nonumber\\
&\stackrel{(b)}{=}\frac{T_{1}^{2}(z^{2}+h^{2})^{-m-3}}{\hat{\mathtt{I}}_{u,v}(z)+\sigma^{2}},
\label{eqn:sinr}  
\end{align}
where $(b)$ follows from \eqref{eqn:intapp1a}.
So, the rate $R$ with the units of bits/s/Hz is given as 
\begin{align}
R&=\log_{2}(1+\gamma(z)),\nonumber\\
&\stackrel{(c)}{=}\log_{2}\bigg(1+\frac{T_{1}^{2}(z^{2}+h^{2})^{-m-3}}{\hat{\mathtt{I}}_{u,v}(z)+\sigma^{2}}\bigg),
\label{eqn:sinrl1r}  
\end{align}
where $(c)$ follows from \eqref{eqn:sinr}. \\ \par
 
We now proceed to understand how to improve the performance of the system by applying the TDMA over the LEDs. 

\section{The TDMA over the LEDs}
\begin{figure}[ht] 
\centering
\resizebox{0.35\textwidth}{!}{%
 \begin{tikzpicture} 
% \node at (1.55,3.75) {\tiny{Generalized top view of the infinite 2D attocell network}};%
% \node at (1.55,3.55) {\tiny{(for a duration of $1/K^2)$.}};%
 %Arrows and marks
 \node at (2.8,0) {\tiny$a$};%H
 \draw[arrows=->](3.25,0)--(3.05,0);
 \draw[arrows=->](2.35,0)--(2.55,0);
 \node at (0.8,0) {\tiny$Ka$};%H
 \draw[arrows=->](1.05,0)--(1.55,0);
 \draw[arrows=->](0.55,0)--(0.05,0);
 \node at (-0.2,1.0) {\tiny$Ka$};%V
 \draw[arrows=->](-0.2,1.25)--(-0.2,1.75);
 \draw[arrows=->](-0.2,0.75)--(-0.2,0.25);
 \node at (3.3,0.5) {\tiny$a$};%V
 \draw[arrows=->](3.3,0.05)--(3.3,0.25);
 \draw[arrows=->](3.3,0.95)--(3.3,0.75);
 %Horizontal dotted
 \draw [dotted,line width= 0.025cm] (-0.4,0.25) -- (3.6,0.25);%
 \draw [dotted,line width= 0.01cm] (-0.4,0.75) -- (3.6,0.75);
 \draw [dotted,line width= 0.01cm] (-0.4,1.25) -- (3.6,1.25);
 \draw [dotted,line width= 0.025cm] (-0.4,1.75) -- (3.6,1.75);%
 \draw [dotted,line width= 0.01cm] (-0.4,2.25) -- (3.6,2.25);
 \draw [dotted,line width= 0.01cm] (-0.4,2.75) -- (3.6,2.75); 
 \draw [dotted,line width= 0.025cm] (-0.4,3.25) -- (3.6,3.25);% 
 %Vertical dotted
 \draw [dotted,line width= 0.025cm] (0.05,-0.1) -- (0.05,3.65);%
 \draw [dotted,line width= 0.01cm] (0.55,-0.1) -- (0.55,3.65);
 \draw [dotted,line width= 0.01cm] (1.05,-0.1) -- (1.05,3.65);
 \draw [dotted,line width= 0.025cm] (1.55,-0.1) -- (1.55,3.65);% 
 \draw [dotted,line width= 0.01cm] (2.05,-0.1) -- (2.05,3.65); 
 \draw [dotted,line width= 0.01cm] (2.55,-0.1) -- (2.55,3.65); 
 \draw [dotted,line width= 0.025cm] (3.05,-0.1) -- (3.05,3.65);% 
 %Row0
 \draw [fill] (-0.2,0) circle [radius=0.05];
 \draw [fill] (1.8,0) circle [radius=0.11];
 %Row1
 \draw [fill] (0.3,0.5) circle [radius=0.05];
 \draw [fill] (0.8,0.5) circle [radius=0.05];
 \draw [fill] (1.3,0.5) circle [radius=0.05];
 \draw [fill] (1.8,0.5) circle [radius=0.05];
 \draw [fill] (2.3,0.5) circle [radius=0.05];
 \draw [fill] (2.8,0.5) circle [radius=0.05];
 %Row2
 \draw [fill] (0.3,1.0) circle [radius=0.05];
 \draw [fill] (0.8,1.0) circle [radius=0.05];
 \draw [fill] (1.3,1.0) circle [radius=0.05];
 \draw [fill] (1.8,1.0) circle [radius=0.05];
 \draw [fill] (2.3,1.0) circle [radius=0.05];
 \draw [fill] (2.8,1.0) circle [radius=0.05];
 %Row3
 \draw [fill] (0.3,1.5) circle [radius=0.11];
 \draw [fill] (0.8,1.5) circle [radius=0.05];
 \draw [fill] (1.3,1.5) circle [radius=0.05];
 \draw [fill] (1.8,1.5) circle [radius=0.11];
 \draw [fill] (2.3,1.5) circle [radius=0.05];
 \draw [fill] (2.8,1.5) circle [radius=0.05];
 \draw [fill] (3.3,1.5) circle [radius=0.11];
 %Row4
 \draw [fill] (-0.2,2.0) circle [radius=0.05];
 \draw [fill] (0.3,2.0) circle [radius=0.05];
 \draw [fill] (0.8,2.0) circle [radius=0.05];
 \draw [fill] (1.3,2.0) circle [radius=0.05];
 \draw [fill] (1.8,2.0) circle [radius=0.05];
 \draw [fill] (2.3,2.0) circle [radius=0.05];
 \draw [fill] (2.8,2.0) circle [radius=0.05];
 \draw [fill] (3.3,2.0) circle [radius=0.05];
 %Row5
 \draw [fill] (-0.2,2.5) circle [radius=0.05];
 \draw [fill] (0.3,2.5) circle [radius=0.05];
 \draw [fill] (0.8,2.5) circle [radius=0.05];
 \draw [fill] (1.3,2.5) circle [radius=0.05];
 \draw [fill] (1.8,2.5) circle [radius=0.05];
 \draw [fill] (2.3,2.5) circle [radius=0.05];
 \draw [fill] (2.8,2.5) circle [radius=0.05];
 \draw [fill] (3.3,2.5) circle [radius=0.05];
 %Row6
 \draw [fill] (-0.2,3) circle [radius=0.05];
 \draw [fill] (0.3,3) circle [radius=0.11];
 \draw [fill] (0.8,3) circle [radius=0.05];
 \draw [fill] (1.3,3) circle [radius=0.05];
 \draw [fill] (1.8,3) circle [radius=0.11];
 \draw [fill] (2.3,3) circle [radius=0.05];
 \draw [fill] (2.8,3) circle [radius=0.05];
 \draw [fill] (3.3,3) circle [radius=0.11];
 %Row7
 \draw [fill] (-0.2,3.5) circle [radius=0.05];
 \draw [fill] (0.3,3.5) circle [radius=0.05];
 \draw [fill] (0.8,3.5) circle [radius=0.05];
 \draw [fill] (1.3,3.5) circle [radius=0.05];
 \draw [fill] (1.8,3.5) circle [radius=0.05];
 \draw [fill] (2.3,3.5) circle [radius=0.05];
 \draw [fill] (2.8,3.5) circle [radius=0.05];
 \draw [fill] (3.3,3.5) circle [radius=0.05];
 \end{tikzpicture}
 }%
\caption{\textit{Generalized top view of the infinite 2D attocell network for a duration of $\frac{1}{K^2}$}. Here, $K=3$, or $K^2=9$ LEDs are symmetrically grouped together into the virtual square groups. Each group is  bordered by boldly dotted lines, and is represented by a collation of $K^2$ smaller square attocells. In the $\frac{1}{K^2}$ duration, only one of the $K^2$ LEDs is LiFi capable and is represented by a larger filled circle. All other non-LiFi LEDs are represented by relatively smaller filled circles. For a given attocell length $a$, the adjacent separation between the LiFi LEDs now becomes $Ka$.}
\label{two_dim1} 
\end{figure}
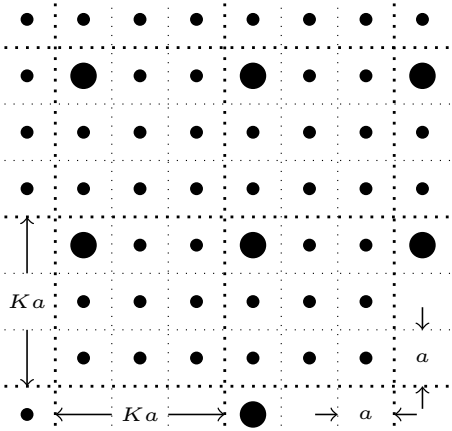

A generalized top view of the infinite 2D plane of Fig.\ref{two_dim} is shown in Fig.\ref{two_dim1}. Out of the infinite number of LEDs over the plane, if every $K^{2}$ LEDs ($K=3$ in Fig.\ref{two_dim1}) are square-symmetrically grouped together with $K$ LEDs on each side of the square, and only one out of those $K^{2}$ LEDs now acts as a LiFi source for a time period of $\frac{1}{K^{2}}$, then the following two things happen. Firstly, in that duration of $\frac{1}{K^2}$, every LiFi source in the infinite plane is separated by a distance of $Ka$ along both the edges of the virtual square; and secondly, the other remaining $K^2-1$ LEDs consecutively take up the LiFi ability over the rest of the duration of $\frac{(K^2-1)}{K^2}$. \par 

Given this time division multiplexing, the PD will experience the signal power from the tagged LED only for a duration of $\frac{1}{K^2}$. Whereas, for the rest of $\frac{K^2-1}{K^2}$ duration, it doesn't experience any. During this time period $\frac{1}{K^2}$, because the separation between every neighbouring LiFi source becomes $Ka$, the expressions in \eqref{eqn:probe} and \eqref{eqn:sinrl1r} modify respectively as  
\begin{align}
P'_{e}=\frac{2(M-1)}{M} Q\Bigg(\frac{\frac{R_{pd}d}{2}-\hat{\mathcal{I}}'_{u,v}(z)}{\sqrt{\sigma^{2}+\hat{\mathtt{I}}'_{u,v}(z)}}\Bigg),
%P_{e}\leq\frac{2(M-1)}{M} Q\Bigg(\frac{\frac{R_{pd}d}{2}-\hat{\mathcal{I}}'_{v}(z)}{\sigma}\Bigg),
\label{eqn:probe1}
\end{align}  
\begin{equation}
R'=\frac{1}{K^2}\log_{2}\bigg(1+\frac{T_{1}^{2}(z^{2}+h^{2})^{-m-3}}{\hat{\mathtt{I}}'_{u,v}(z)+\sigma^{2}}\bigg),
%R=\frac{1}{K}\log_{2}\bigg(1+\frac{h^{-2m-6}}{\hat {\mathtt{I}}'_{v}+\Omega}\bigg).
\label{eqn:sinrl1r1}  
\end{equation}
where $\hat{\mathcal{I}}'_{u,v}(z)=\hat{\mathcal{I}}_{u,v}(z)|_{a\rightarrow Ka}$ and $\hat{\mathtt{I}}'_{u,v}(z)=\hat{\mathtt{I}}_{u,v}(z)|_{a\rightarrow Ka}$.

Using the modified expressions, \eqref{eqn:probe1} and \eqref{eqn:sinrl1r1}, relative to this TDMA over the LEDs, an exact expression for the goodput $G$ of the system can now be expressed as 
\begin{align}
G&=R'\times(1-P'_{e}),\nonumber\\
&=\frac{1}{K^2}\log_{2}\bigg(1+\frac{T_{1}^{2}(z^{2}+h^{2})^{-m-3}}{\hat{\mathtt{I}}'_{u,v}(z)+\sigma^{2}}\bigg)\nonumber\\
&\ \ \ \ \ \ \ \ \ \ \times\Bigg(1-\frac{2(M-1)}{M} Q\Bigg(\frac{\frac{R_{pd}d}{2}-\hat{\mathcal{I}}'_{u,v}(z)}{\sqrt{\sigma^{2}+\hat{\mathtt{I}}'_{u,v}(z)}}\Bigg)\Bigg).
\label{eqn:good}
\end{align}
%This expression, as a matter of fact, is same for both one and two dimensional arrangements, save the fact for the trivial changes that have to occur in $\mathtt{I}'_{k}(z)$ and $\mathcal{I}'_{k}(z)$.
\subsection{Optimal time scheduling}
It is of significant importance to exploit this time scheduling to mitigate the effect of interference and maximize the system performance. Evidently, as the TDMA parameter $K$ increases, the LiFi LEDs are separated by a distance of $Ka$ and therefore the co-channel interference also considerably reduces. By analyzing the modified terms $\hat{\mathcal{I}}'_{u,v}(z)$ and $\hat{\mathtt{I}}'_{u,v}(z)$, it can be seen that when $K$ gets bigger and for $\beta>2$,
%\begin{align*}
 \[\mathbb{K}_{\frac{\beta}{2}-1}\bigg(\frac{2\pi h\sqrt{f^2+w^2}}{Ka}\bigg) \ll \mathbb{K}_{\beta-1}\bigg(\frac{2\pi h\sqrt{f^2+w^2}}{Ka}\bigg). \] %, \nonumber\\
%\iff & \hat{\mathcal{I}}'_{u,v}(z) \ll \hat{\mathtt{I}}'_{u,v}(z).
%\end{align*} 
This means the denominator inside the $Q$ function in \eqref{eqn:probe1} reduces more rapidly than the numerator, implying that the probability of error $P'_{e}$ dwindles rapidly over this increase in $K$. Moreover, the rate $R'$ is also bound to increase due to a decrease in interference but only upto a certain value $K$. The reason lies in the fact that $K^2$, which resides as a denominator in the expression for $R'$, dominates the logarithmic behaviour of the rest of the expression. This work essentially focusses on maximizing the goodput $G$ because it represents a joint optimization between the $P'_{e}$ and the rate $R'$ of the time scheduled system. That is to say, the system must be optimally scheduled such that the $P'_{e}$ is relatively minimized upto the extent that the rate $R'$ is maximized. The optimum value of $K$ is given as 
\[ K^{*}=\arg\max \limits_{K}G. \]
This value can easily be evaluated for its existence by performing numerical simulations over the characterization for $G$ using \eqref{eqn:good}. However, it is to note that the $K$ at which $R'$ may become maximum may not be equal to $K^{*}$. This discussion can be further understood with the following numerical simulations.
 
\subsection{Numerical analysis} 
All numerical simulations have been performed using the parameters in Table II. 
\begin{table}
\caption{Parameters considered for numerical simulations}
\label{abc}
\begin{tabular}{ | m{3cm} | m{1cm}| m{2cm} | m{1.0cm} |} 
\hline
Parameter& Symbol & Value & Unit \\ 
\hline
Temperature of Operation & $T$ & $300$ & K \\ 
\hline
Noise power spectral density at Photodiode & $N_{o}$ & $4.14\times10^{-21}$ & A$^{2}$Hz$^{-1}$ \\
\hline
Modulation bandwidth of LED & $W$ & $40\times10^{6}$ & Hz \\
\hline
Area of Photodiode & $A_{pd}$ & $10^{-4}$ & m$^{2}$ \\
\hline
Responsivity of PD & $R_{pd}$ & $0.1$ & AW$^{-1}$ \\
\hline
Order of the PAM & $M$ & $8$ & \\
\hline
Optical power constant & $A$ & $1$ & W\\
\hline
Field Of View of PD & $\theta_{f}$ & $90$ & degrees \\
\hline
Half power semi angle of the LEDs & $\theta_{h}$ & $60$ & degrees \\
\hline 
\end{tabular}
\end{table}

\begin{figure}[ht]
\centering
\definecolor{mycolor1}{rgb}{0.00000,0.44700,0.74100}%
\definecolor{mycolor2}{rgb}{0.85000,0.32500,0.09800}%
\definecolor{mycolor3}{rgb}{0.92900,0.69400,0.12500}%
\begin{tikzpicture}

\begin{axis}[%
width=7cm,
height=5cm,
scale only axis,
xmin=1,
xmax=15,
xlabel style={font=\color{white!15!black}},
xlabel={TDMA Parameter $(K)$},
ymin=0.1,
ymax=1,
ylabel style={font=\color{white!15!black}},
ylabel={Probability of Error $(P'_{e})$},
axis background/.style={fill=white},
xmajorgrids,
ymajorgrids,
legend style={font=\fontsize{7}{5}\selectfont,at={(0.5190,0.548)}, anchor=south west, legend cell align=left, align=left, draw=white!15!black}
] 
\addplot [color=black, line width=1.5pt]
  table[row sep=crcr]{%
1	1\\
2	0.999999999379809\\
3	0.99746581\\
4	0.883754\\
5	0.611999\\
6	0.398017\\
7	0.276795\\
8	0.211063\\
9	0.174003\\
10	0.151925\\
11	0.138055\\
12	0.128905\\
13	0.122577\\
14	0.117975\\
15	0.114439\\
16	0.111553\\
17	0.10905\\
18	0.106751\\
19	0.104541\\
20	0.102345\\
};
\addlegendentry{Analytical $h/a=3$}

\addplot [color=black, dashed, line width=1.5pt]
  table[row sep=crcr]{%
1	1\\
2	0.9999999882392\\
3	0.99784126\\
4	0.9325631\\
5	0.757697\\
6	0.578318\\
7	0.448727\\
8	0.364716\\
9	0.31115\\
10	0.276445\\
11	0.253371\\
12	0.237602\\
13	0.226532\\
14	0.218554\\
15	0.212647\\
16	0.208143\\
17	0.204594\\
18	0.201692\\
19	0.19922\\
20	0.197025\\
};
\addlegendentry{Analytical $h/a=5$}

\addplot [color=black, dotted, line width=1.5pt]
  table[row sep=crcr]{%
1	1\\
2	0.999999834991\\
3	0.99715524\\
4	0.9447929\\
5	0.816244\\
6	0.673764\\
7	0.557711\\
8	0.473609\\
9	0.414946\\
10	0.37422\\
11	0.345683\\
12	0.325389\\
13	0.310717\\
14	0.299931\\
15	0.291869\\
16	0.28574\\
17	0.280998\\
18	0.277257\\
19	0.274243\\
20	0.271752\\
};
\addlegendentry{Analytical $h/a=7$}

\addplot [color=black, draw=none, mark size=4pt, mark=star, mark options={solid, black}]
  table[row sep=crcr]{%
1	1\\
2	0.999999999379809\\
3	0.99746581\\
4	0.883754\\
5	0.611999\\
6	0.398017\\
7	0.276795\\
8	0.211063\\
9	0.174003\\
10	0.151925\\
11	0.138055\\
12	0.128905\\
13	0.122577\\
14	0.117975\\
15	0.114439\\
16	0.111553\\
17	0.10905\\
18	0.106751\\
19	0.104541\\
20	0.102345\\
};
\addlegendentry{Numerical $h/a=3$}

\addplot [color=black, draw=none, mark size=4pt, mark=o, mark options={solid, black}]
  table[row sep=crcr]{%
1	1\\
2	0.9999999882392\\
3	0.99784126\\
4	0.9325631\\
5	0.757697\\
6	0.578318\\
7	0.448727\\
8	0.364716\\
9	0.31115\\
10	0.276445\\
11	0.253371\\
12	0.237602\\
13	0.226532\\
14	0.218554\\
15	0.212647\\
16	0.208143\\
17	0.204594\\
18	0.201692\\
19	0.19922\\
20	0.197025\\
};
\addlegendentry{Numerical $h/a=5$}

\addplot [color=black, draw=none, mark size=4pt, mark=square, mark options={solid, black}]
  table[row sep=crcr]{%
1	1\\
2	0.999999834991\\
3	0.99715524\\
4	0.9447929\\
5	0.816244\\
6	0.673764\\
7	0.557711\\
8	0.473609\\
9	0.414946\\
10	0.37422\\
11	0.345683\\
12	0.325389\\
13	0.310717\\
14	0.299931\\
15	0.291869\\
16	0.28574\\
17	0.280998\\
18	0.277257\\
19	0.274243\\
20	0.271752\\
};
\addlegendentry{Numerical $h/a=7$}

\end{axis}
\end{tikzpicture}%
\caption{The variation of probability of error $P'_{e}$, for both numerical and analytical simulations, is plotted against the TDMA parameter $K$ for different values of the ratio $h/a$ of LED installation. Here, the PD is assumed to be located at $z_x=z_y=0$.} 
\label{pe1}
\end{figure}
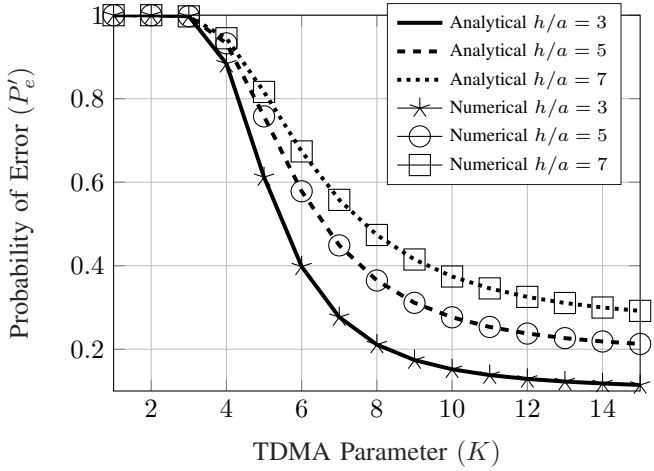

The LEDs are  assumed to be installed at different values of height to LED separation ratio $h/a\in\{3,5,7\}$. From \cite{atchu2}, it is sufficient to consider the interference summation order of $u=v=2$, which itself is a good approximation over the considered values of $K\in\{1, 2, 3, ... , 15\}$. First of all, the behaviour of $P'_{e}$ and $R'$ against $K$ is observed in Fig.\ref{pe1} and \ref{rate1}. The graph for goodput is then drawn in Fig.\ref{good1}. 
\subsubsection{Probability of error $P'_{e}$ (Fig.\ref{pe1})} When the value of $K$ increases, the probability of error reduces rapidly. For both numerical and analytical simulations over different values of $h/a$. Importantly, the numerical simulations performed without any of the approximations made in this paper tightly bound with the analytical ones, which actually validates the latter. The drop in $P'_{e}$ generally starts after $K>2$ and for any given $K$, the $P'_{e}$ is always lesser for a lesser value of $h/a$. 

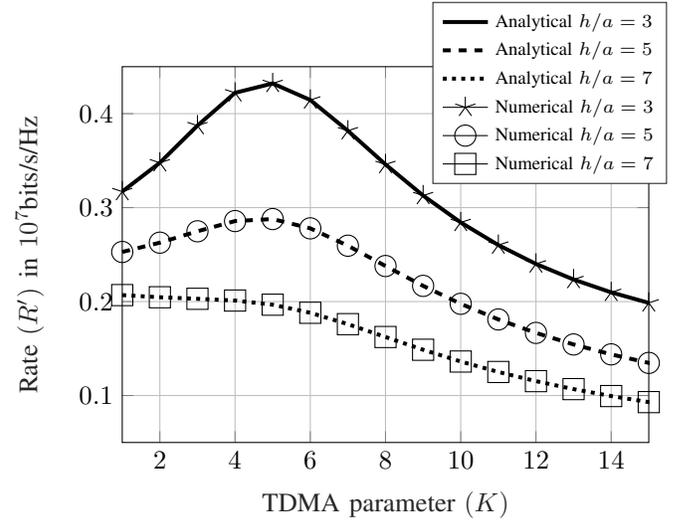
\begin{figure}[ht]
\centering
\definecolor{mycolor1}{rgb}{0.00000,0.44700,0.74100}%
\definecolor{mycolor2}{rgb}{0.85000,0.32500,0.09800}%
\definecolor{mycolor3}{rgb}{0.92900,0.69400,0.12500}%
\begin{tikzpicture}

\begin{axis}[%
width=7cm,
height=5cm,
scale only axis,
xmin=1,
xmax=15,
xlabel style={font=\color{white!15!black}},
xlabel={TDMA parameter $(K)$},
ymin=0.05,
ymax=0.45,
ylabel style={font=\color{white!15!black}},
ylabel={Rate $(R')$ in $10^{7}$bits/s/Hz},
axis background/.style={fill=white},
xmajorgrids,
ymajorgrids,
legend style={font=\fontsize{7}{5}\selectfont,at={(0.5890,0.688)}, anchor=south west, legend cell align=left, align=left, draw=white!15!black}
]
\addplot [color=black, line width=1.5pt]
  table[row sep=crcr]{%
1	0.317072\\
2	0.347766\\
3	0.387296\\
4	0.422207\\
5	0.432167\\
6	0.414547\\
7	0.381821\\
8	0.345952\\
9	0.312827\\
10	0.284208\\
11	0.260127\\
12	0.24008\\
13	0.223484\\
14	0.209807\\
15	0.198605\\
16	0.189522\\
17	0.182276\\
18	0.176657\\
19	0.172515\\
20	0.169764\\
};
\addlegendentry{Analytical $h/a=3$}

\addplot [color=black, dashed, line width=1.5pt]
  table[row sep=crcr]{%
1	0.2528\\
2	0.262695\\
3	0.274805\\
4	0.285691\\
5	0.28783\\
6	0.277926\\
7	0.259469\\
8	0.237851\\
9	0.216707\\
10	0.19762\\
11	0.180973\\
12	0.166644\\
13	0.154352\\
14	0.143799\\
15	0.134719\\
16	0.126883\\
17	0.120104\\
18	0.114226\\
19	0.10912\\
20	0.104678\\
};
\addlegendentry{Analytical $h/a=5$}

\addplot [color=black, dotted, line width=1.5pt]
  table[row sep=crcr]{%
1	0.206884\\
2	0.204599\\
3	0.202958\\
4	0.201091\\
5	0.196676\\
6	0.188071\\
7	0.175959\\
8	0.162276\\
9	0.14871\\
10	0.136192\\
11	0.125054\\
12	0.115305\\
13	0.106818\\
14	0.0994281\\
15	0.0929758\\
16	0.0873206\\
17	0.0823435\\
18	0.0779455\\
19	0.0740443\\
20	0.0705713\\
};
\addlegendentry{Analytical $h/a=7$}

\addplot [color=black, draw=none, mark size=4pt, mark=star, mark options={solid, black}]
  table[row sep=crcr]{%
1	0.317072\\
2	0.347766\\
3	0.387296\\
4	0.422207\\
5	0.432167\\
6	0.414547\\
7	0.381821\\
8	0.345952\\
9	0.312827\\
10	0.284208\\
11	0.260127\\
12	0.24008\\
13	0.223484\\
14	0.209807\\
15	0.198605\\
16	0.189522\\
17	0.182276\\
18	0.176657\\
19	0.172515\\
20	0.169764\\
};
\addlegendentry{Numerical $h/a=3$}

\addplot [color=black, draw=none, mark size=4pt, mark=o, mark options={solid, black}]
  table[row sep=crcr]{%
1	0.2528\\
2	0.262695\\
3	0.274805\\
4	0.285691\\
5	0.28783\\
6	0.277926\\
7	0.259469\\
8	0.237851\\
9	0.216707\\
10	0.19762\\
11	0.180973\\
12	0.166644\\
13	0.154352\\
14	0.143799\\
15	0.134719\\
16	0.126883\\
17	0.120104\\
18	0.114226\\
19	0.10912\\
20	0.104678\\
};
\addlegendentry{Numerical $h/a=5$}

\addplot [color=black, draw=none, mark size=4pt, mark=square, mark options={solid, black}]
  table[row sep=crcr]{%
1	0.206884\\
2	0.204599\\
3	0.202958\\
4	0.201091\\
5	0.196676\\
6	0.188071\\
7	0.175959\\
8	0.162276\\
9	0.14871\\
10	0.136192\\
11	0.125054\\
12	0.115305\\
13	0.106818\\
14	0.0994281\\
15	0.0929758\\
16	0.0873206\\
17	0.0823435\\
18	0.0779455\\
19	0.0740443\\
20	0.0705713\\
};
\addlegendentry{Numerical $h/a=7$}

\end{axis}
\end{tikzpicture}%
\caption{The variation of rate $R'$, for both numerical and analytical simulations, is plotted against the TDMA parameter $K$ for different values of the ratio $h/a$ of LED installation. Here, the PD is assumed to be located at $z_x=z_y=0$.}
\label{rate1}
\end{figure}

\subsubsection{Rate $R'$ (Fig.\ref{rate1})} As mentioned earlier, it can be seen that the rate $R'$ attains a peak value at a particular value of $K$. It improves till the peak and decreases thereafter, which clearly implies the effect of a logarithmic increase due to decrease in interference before the peak, and the dominant effect of $K^2$ in the denominator of \eqref{eqn:sinrl1r1} after the peak. Also, the numerical simulations performed without any of the approximations made in this paper tightly bound with the analytical ones, which actually validates the latter. Parallelly, the rate is always higher for a lower value of $h/a$ for any given value of $K$.

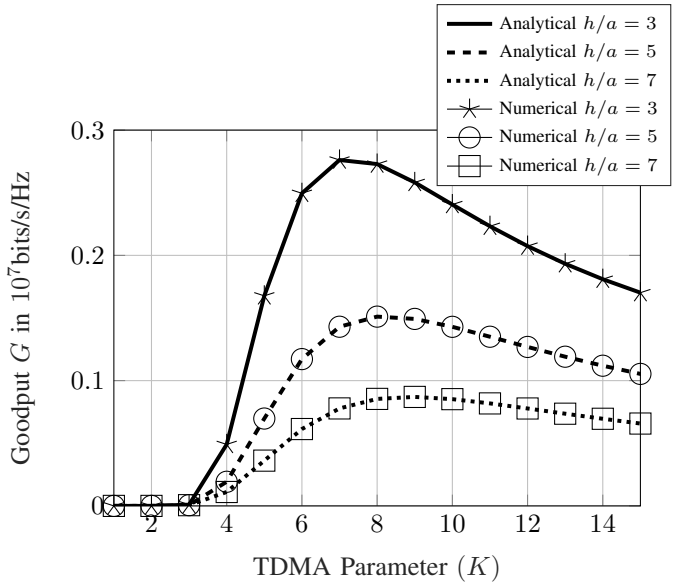
\begin{figure}[ht]
\centering
\definecolor{mycolor1}{rgb}{0.00000,0.44700,0.74100}%
\definecolor{mycolor2}{rgb}{0.85000,0.32500,0.09800}%
\definecolor{mycolor3}{rgb}{0.92900,0.69400,0.12500}%
\begin{tikzpicture}

\begin{axis}[%
width=7cm,
height=5cm,
scale only axis,
xmin=1,
xmax=15,
xlabel style={font=\color{white!15!black}},
xlabel={TDMA Parameter $(K)$},
ymin=0,
ymax=0.3,
ylabel style={font=\color{white!15!black}},
ylabel={Goodput $G$ in $10^{7}$bits/s/Hz},
axis background/.style={fill=white},
xmajorgrids,
ymajorgrids,
legend style={font=\fontsize{7}{5}\selectfont,at={(0.6130,0.851)}, anchor=south west, legend cell align=left, align=left, draw=white!15!black}
]
\addplot [color=black, line width=1.5pt]
  table[row sep=crcr]{%
1	0\\
2	2.15681e-10\\
3	0.000981481\\
4	0.0490799\\
5	0.167682\\
6	0.249549\\
7	0.276126\\
8	0.272886\\
9	0.258217\\
10	0.240566\\
11	0.223243\\
12	0.207391\\
13	0.193305\\
14	0.18096\\
15	0.170221\\
16	0.160927\\
17	0.15292\\
18	0.146054\\
19	0.140202\\
20	0.135255\\
};
\addlegendentry{Analytical $h/a=3$}

\addplot [color=black, dashed, line width=1.5pt]
  table[row sep=crcr]{%
1	0\\
2	3.08951e-09\\
3	0.000593232\\
4	0.0192661\\
5	0.0697421\\
6	0.117197\\
7	0.143038\\
8	0.151102\\
9	0.149272\\
10	0.142963\\
11	0.135052\\
12	0.126902\\
13	0.119111\\
14	0.111914\\
15	0.105373\\
16	0.0994791\\
17	0.0941907\\
18	0.0894558\\
19	0.0852212\\
20	0.0814359\\
};
\addlegendentry{Analytical $h/a=5$}

\addplot [color=black, dotted, line width=1.5pt]
  table[row sep=crcr]{%
1	0\\
2	3.37606e-08\\
3	0.000577365\\
4	0.0111017\\
5	0.0361403\\
6	0.0613554\\
7	0.0778247\\
8	0.0854208\\
9	0.0870032\\
10	0.0852248\\
11	0.0818206\\
12	0.077774\\
13	0.0736008\\
14	0.0695547\\
15	0.0657501\\
16	0.0622298\\
17	0.0589999\\
18	0.0560497\\
19	0.0533605\\
20	0.0509111\\
};
\addlegendentry{Analytical $h/a=7$}

\addplot [color=black, draw=none, mark size=4pt, mark=star, mark options={solid, black}]
  table[row sep=crcr]{%
1	0\\
2	2.15681e-10\\
3	0.000981481\\
4	0.0490799\\
5	0.167682\\
6	0.249549\\
7	0.276126\\
8	0.272886\\
9	0.258217\\
10	0.240566\\
11	0.223243\\
12	0.207391\\
13	0.193305\\
14	0.18096\\
15	0.170221\\
16	0.160927\\
17	0.15292\\
18	0.146054\\
19	0.140202\\
20	0.135255\\
};
\addlegendentry{Numerical $h/a=3$}

\addplot [color=black, draw=none, mark size=4pt, mark=o, mark options={solid, black}]
  table[row sep=crcr]{%
1	0\\
2	3.08951e-09\\
3	0.000593232\\
4	0.0192661\\
5	0.0697421\\
6	0.117197\\
7	0.143038\\
8	0.151102\\
9	0.149272\\
10	0.142963\\
11	0.135052\\
12	0.126902\\
13	0.119111\\
14	0.111914\\
15	0.105373\\
16	0.0994791\\
17	0.0941907\\
18	0.0894558\\
19	0.0852212\\
20	0.0814359\\
};
\addlegendentry{Numerical $h/a=5$}

\addplot [color=black, draw=none, mark size=4pt, mark=square, mark options={solid, black}]
  table[row sep=crcr]{%
1	0\\
2	3.37606e-08\\
3	0.000577365\\
4	0.0111017\\
5	0.0361403\\
6	0.0613554\\
7	0.0778247\\
8	0.0854208\\
9	0.0870032\\
10	0.0852248\\
11	0.0818206\\
12	0.077774\\
13	0.0736008\\
14	0.0695547\\
15	0.0657501\\
16	0.0622298\\
17	0.0589999\\
18	0.0560497\\
19	0.0533605\\
20	0.0509111\\
};
\addlegendentry{Numerical $h/a=7$}

\end{axis}
\end{tikzpicture}%
\caption{The variation of Goodput $G=R'\times(1-P'_{e})$, for both numerical and analytical simulations, is plotted against the TDMA parameter $K$ for different values of the ratio $h/a$ of LED installation. Here, the PD is assumed to be located at $z_x=z_y=0$.}
\label{good1}
\end{figure}

\subsubsection{Goodput $G$ (Fig.\ref{good1})} It was observed in Fig.\ref{pe1} and \ref{rate1} that though the rate $R'$ may have been maximized at a particular value of $K$, the $P'_{e}$ was still large enough to cause a decrease in the system performance. For example, at $h/a=3$, and at $K=5$ where the rate was maximized to $R'=0.43\times10^{7}$bits/s/Hz, the value of $P'_{e}=0.6$ was still large enough, and it was desirable to achieve a lower value. So, it is essential to perform a trade-off between the allowable probability of error and the rate that can be achieved; and this is taken care of by the goodput $G$ of the system. The value of $K^{*}(=\arg\max \limits_{K}G)$ is higher than that obtained to maximize the rate. It occurs between $K=6$ and $8$ for different values of $h/a$ considered. Here too, the numerical simulations performed without any of the approximations made in this paper tightly bound with the analytical ones, which actually validates the latter. Also, $G$ is always higher for a lower value of $h/a$ at any given $K$. 

Above all, the fact is clearly established that co-channel interference can be mitigated and the performance of a LiFi attocell network can be improved with TDMA over the LEDs, and there exists an optimal value of $K$ at which the goodput $G$, an evidently appropriate parameter, can be maximized. 
      
\section{Conclusion}
This work uses the interference approximation in \cite{atchu2} to provide closed form expressions for the goodput $G$ of a time scheduled LiFi system in its downlink. It was shown through analytical simulations that the co-channel interference can be mitigated and the performance of a LiFi attocell network can be improved with TDMA over the LEDs. Moreover, the existence of an optimum scheduling parameter $K$ was shown at which $G$ was maximized. 
\bibliographystyle{IEEEtran}
\bibliography{IEEEfull,referencetdm.bib}

\end{document}